 \documentclass[final,5p,times,twocolumn]{elsarticle}

\usepackage{amssymb}
\usepackage{amsthm}
\usepackage{amsmath}
\usepackage{upgreek}
\usepackage{csquotes}

\usepackage{lineno}
\makeatletter
\g@addto@macro\endfrontmatter{\enlargethispage{-2\baselineskip}}
\makeatother

\usepackage{etoolbox}
\makeatletter
    \patchcmd{\@author}{\global\let\@fnmark\@empty}{\global\let\@fnmark\@empty\global\let\@corref\@empty}{}{\@latex@error{Failed to patch \string\@author for \string\@corref reset}}
\makeatother

\usepackage{tabulary}
\usepackage{array}
\usepackage{booktabs}
\usepackage{tabularx}
\newcolumntype{Y}{>{\centering\arraybackslash}X}
\newcolumntype{L}{>{\raggedright\arraybackslash}X}
\newcolumntype{R}{>{\raggedleft\arraybackslash}X}
\newcolumntype{a}{>{\hsize=.2\hsize}X}
\newcolumntype{b}{>{\hsize=1.8\hsize}X}
\newcolumntype{t}{>{\hsize=1.5\hsize}X}

\usepackage{gensymb}

\usepackage{hyperref}

\usepackage{siunitx}
\usepackage{tikz}
\usepackage{circuitikz}

\usepackage{subfig}

\newcommand*{\doi}[1]{\href{http://dx.doi.org/#1}{doi: #1}}

 \journal{Nuclear Inst. and Methods in Physics Research A}
\begin{document}

\begin{frontmatter}

%% Title, authors and addresses

%% use the tnoteref command within \title for footnotes;
%% use the tnotetext command for theassociated footnote;
%% use the fnref command within \author or \address for footnotes;
%% use the fntext command for theassociated footnote;
%% use the corref command within \author for corresponding author footnotes;
%% use the cortext command for theassociated footnote;
%% use the ead command for the email address,
%% and the form \ead[url] for the home page:
%% \title{Title\tnoteref{label1}}
%% \tnotetext[label1]{}
%% \author{Name\corref{cor1}\fnref{label2}}
%% \ead{email address}
%% \ead[url]{home page}
%% \fntext[label2]{}
%% \cortext[cor1]{}
%% \address{Address\fnref{label3}}
%% \fntext[label3]{}

\title{Process Quality Control Strategy for the Phase-2 Upgrade of the CMS Outer Tracker and High Granularity Calorimeter}

%% use optional labels to link authors explicitly to addresses:
%% \author[label1,label2]{}
%% \address[label1]{}
%% \address[label2]{}

\author{V. Hinger\corref{cor1}}
\ead{viktoria.hinger@oeaw.ac.at}
\author{on behalf of the CMS Collaboration}

\cortext[cor1]{Corresponding author}

\address{Institute of High Energy Physics, Austrian Academy of Sciences, Nikolsdorfer Gasse 18, 1050 Wien, Austria}

\begin{abstract}
%% Text of abstract
 Between 2025 and 2027, some essential components of the CMS (Compact Muon Solenoid) detector -- most notably the tracker and the calorimeter endcap -- will be upgraded to prepare for HL-LHC (High Luminosity Large Hadron Collider) conditions.
 The upgraded CMS Outer Tracker and parts of the new CMS High Granularity Calorimeter (\mbox{HGCAL}) will encompass over $50{,}000$ new silicon sensors covering a total area of about $800\,\si{\metre\squared}$. The sensor series production requires a dedicated strategy to monitor the quality and stability of the production process. The strategy is based on a test structure set that provides quick and easy access to critical process parameters. These include parameters not directly accessible on the sensors (e.g. oxide charge concentration and interface trap density) and parameters requiring potentially destructive measurements (e.g. dielectric strength). The set is implemented at least twice on each production wafer. It is divided into test structures for initial evaluation of the most relevant process parameters and structures for in-depth analysis. All structures can be contacted using a 20-needle probe card and an automated positioning stage. With this system, the initial analysis of one wafer is possible in about 30 minutes.
 In this paper, the CMS collaboration presents the quality assurance plan for the Phase-2 Upgrade with a focus on process quality control. We cover sensor process specifics, the layout of the test structure set that will be implemented in the production runs for the CMS Outer Tracker and \mbox{HGCAL}, and measurement results illustrating the functionality of the included test structures.
\end{abstract}

\begin{keyword}
%% keywords here, in the form: keyword \sep keyword
silicon sensors \sep CMS tracker \sep CMS \mbox{HGCAL} \sep test structures
%% PACS codes here, in the form: \PACS code \sep code

%% MSC codes here, in the form: \MSC code \sep code
%% or \MSC[2008] code \sep code (2000 is the default)

\end{keyword}

\end{frontmatter}

%% main text
\section{Introduction}
\label{Intro}
	
	During the Long Shutdown 3, between 2025 and 2027, the CERN Large Hadron Collider (LHC) will undergo a major upgrade to allow the instantaneous luminosity to reach $7.5\times10^{34}\,\si{\per\square\centi\metre\per\second}$, corresponding to an increase of the collision rate by a factor of 5 compared to the present. By 2037, the integrated luminosity will have increased tenfold, reaching $3000\,\si{\per\femto\barn}$ ($4000\,\si{\per\femto\barn}$ for the ultimate scenario). These high-luminosity conditions make it necessary to replace some existing components of the CMS (Compact Muon Solenoid) detector~\cite{Collaboration_2008} -- most notably the CMS Tracker~\cite{CMSTracker} and the CMS Endcap Calorimeter~\cite{CMSHGCAL} -- as part of the CMS Phase-2 Upgrade.
	
	Extensive studies have demonstrated the robustness of p-type silicon and made it a preferred detector base line material for high-radiation environments~\cite{Adam_2017,CURRAS201760}. Silicon sensors will constitute the full volume of the CMS tracker and parts of the new High Granularity Calorimeter (\mbox{HGCAL}). For these detector parts, over 50,000 silicon sensors will be manufactured between 2020 and 2023. This large-scale sensor series production requires a dedicated quality assurance plan. The strategy described in this paper relies on three main quality control mechanisms: sensor quality control, process quality control, and irradiation tests.
	
	Process quality control tracks the quality and stability of the manufacturing process throughout production. For this purpose, critical process parameters are monitored on test structures.
	CMS designed a set of test structures that facilitate a quick assessment of the most important process parameters and provide the means for in-depth analysis in case of problems.
	This set is the basis for automated process quality control during the upcoming silicon sensor series production for the CMS Phase-2 Upgrade.

\section{Silicon sensors for the CMS Phase-2 Upgrade}

\subsection{Outer Tracker}
\label{Rint}
	
	\begin{figure}[tb]
		\centering
		\subfloat[DC-coupled macro-pixel sensor layout with punch-through bias.]{
			\includegraphics[width=\linewidth]{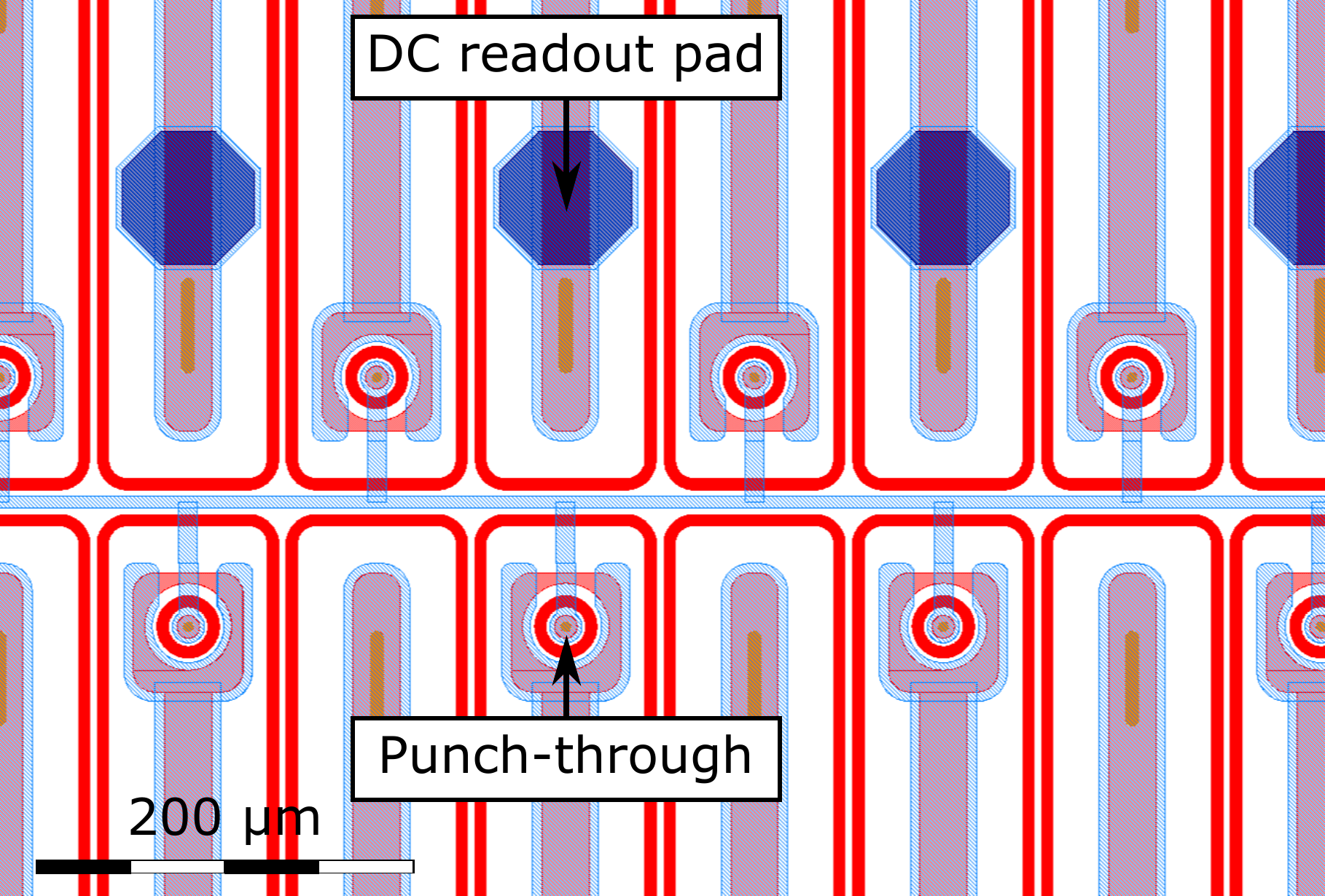}}%
		\newline
%	\hfill
		\subfloat[AC-coupled strip sensor layout with polysilicon bias resistors.]{
			\includegraphics[width=\linewidth]{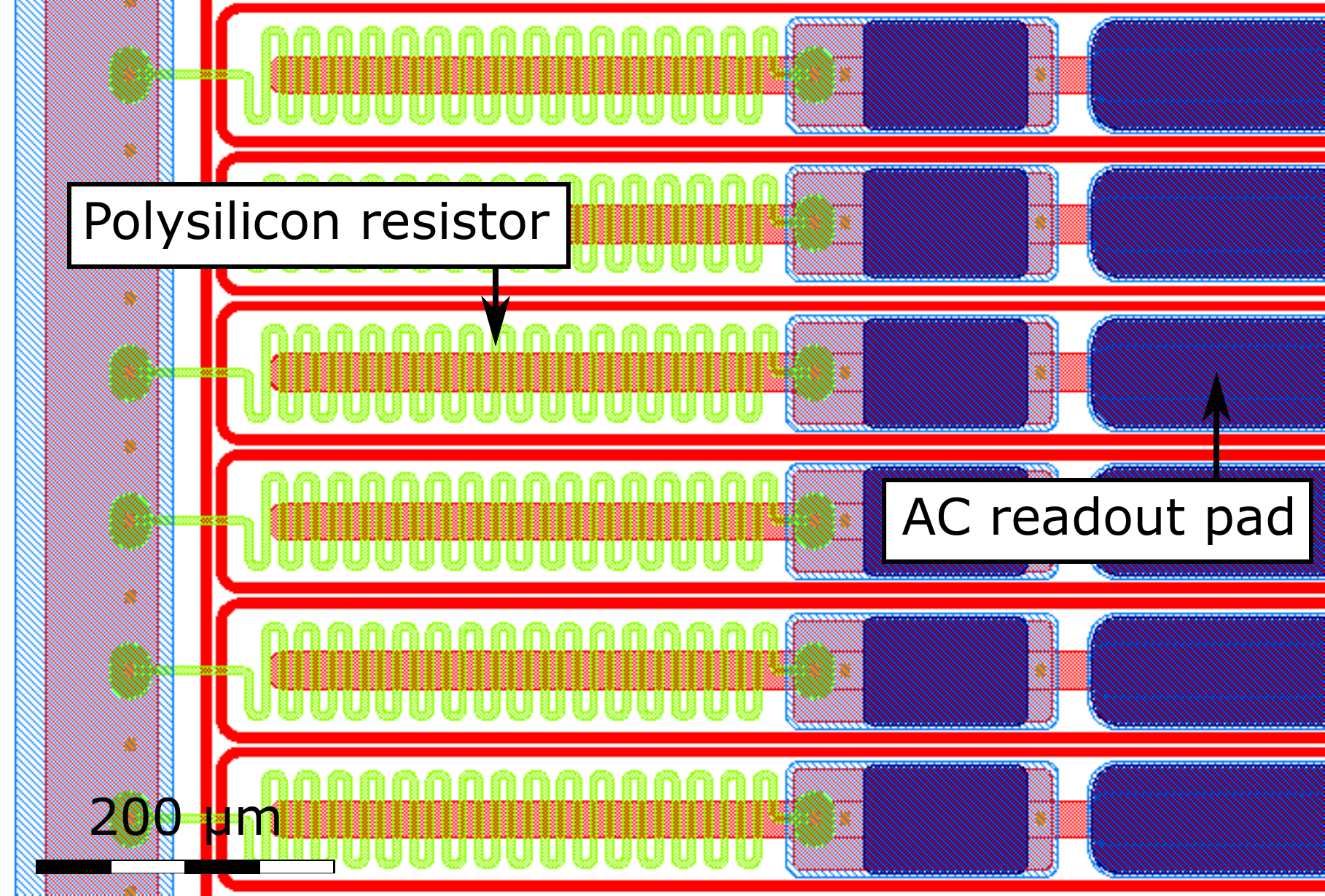}}%
		%\par\smallskip
		\caption{Layout close-up of the outer tracker macro-pixel and strip sensors, illustrating the differences between DC-coupled (a) and AC-coupled (b) processes.}
		\label{fig:TrackerSensors}
	\end{figure}
	
	The upgraded Outer Tracker of the CMS detector will encompass about $200\,\si{\metre\squared}$ of silicon sensors, amounting to about 24,000 wafers. Each tracker module consists of a pair of stacked sensors. This design enables discrimination of the particle track transverse momentum ($p_{\text{T}}$) at module level. Two types of modules exist depending on the location in the tracker. In the inner part of the Outer Tracker (i.e. closer to the beam line), the modules consist of a macro-\textbf{p}ixel sensor and a single-sided \textbf{s}trip sensor (\textbf{PS} modules). And the outer part of the tracker houses modules consisting of \textbf{2 s}trip sensors (\textbf{2S} modules).
	All sensors are n-in-p type produced on $6\,''$ wafers, but feature different biasing and read-out schemes. The strip sensors are AC-coupled and biased via polysilicon resistors, while the macro-pixel sensors are DC-coupled and additionally feature punch-through structures for biasing during testing (Figure~\ref{fig:TrackerSensors}). The production of AC-coupled strip sensors requires more steps and thus a different process than the production of DC-coupled pixels.
	For a detailed discussion of the CMS Outer Tracker Phase-2 Upgrade, see~\cite{HSTD12CMSOT}.

\subsection{High Granularity Calorimeter}

	\begin{figure}[tb]
		\centering
		\includegraphics[width=\linewidth]{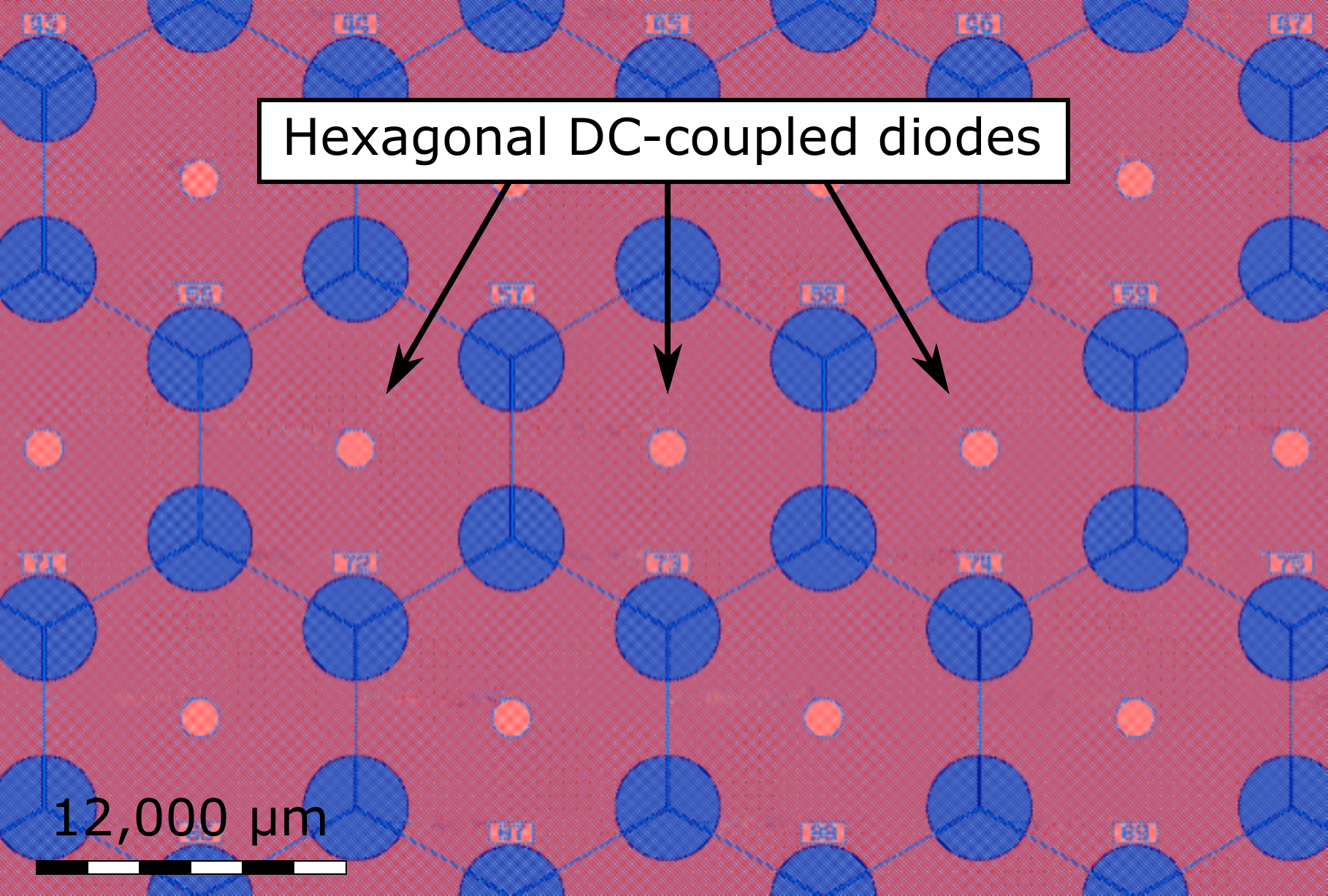}
		\caption{Layout close-up of a HGCAL sensor with hexagonal planar DC-coupled diodes as active elements.}
		\label{fig:HGCAL}
	\end{figure}
	
	CMS will build the new calorimeter endcap (CE) as a sampling calorimeter with unprecedented transverse and longitudinal granularity. For this reason, the CE is also called the High Granularity Calorimeter (\mbox{HGCAL}). The full electromagnetic compartment (\mbox{CE-E}) and the high-fluence regions of the hadronic compartment (\mbox{CE-H}) will utilize silicon sensors as active detector material. The sensors are hexagonal and use n-in-p technology on $8\,''$ wafers. The active elements on the silicon sensors are hexagonal planar DC-coupled diodes (Figure~\ref{fig:HGCAL}). In total, the CE will encompass about $600\,\si{\metre\squared}$ of silicon sensors, amounting to about 28,000 wafers\footnote{Preliminary number. The total number of sensors differs. Some wafers may house two smaller sensors for optimized coverage of the calorimeter volume.}.
	
	The $8\,''$ process is new for large-area sensors in high-energy physics applications and differs substantially from the well-known $6\,''$ process used for the CMS Tracker. Furthermore, the calorimeter sensors will come in three different wafer thicknesses ($300\,\si{\micro\metre}$, $200\,\si{\micro\metre}$, and $120\,\si{\micro\metre}$), accounting for regions of different fluence in the calorimeter volume. In contrast to the $300\,\si{\micro\metre}$ and $200\,\si{\micro\metre}$ thick wafers, which use a standard float zone process, the $120\,\si{\micro\metre}$ thick wafers use an epitaxial production process.
	For further information about the \mbox{HGCAL} silicon sensors, see~\cite{HSTD12HGCAL}.

\section{Silicon sensor quality assurance plan}

CMS has developed a plan to ensure that all sensors that will be integrated in the detector meet predefined quality standards. Firstly, all sensors are pretested by the vendor. Only the sensors that meet the specifications and all corresponding test structures are distributed to the test centers of the sensor working group. Subsequently, the test centers perform three main quality control procedures:
\begin{enumerate}
	\item \emph{Sensor quality control (SQC):} About $10\,\%$ of sensors of each delivered batch are characterized to ensure that they fully satisfy the specifications.
	\item \emph{Process quality control (PQC):} The quality and stability of the production process are tracked by measuring process parameters on test structures. At least $20\,\%$ of the delivered wafers are characterized.
	\item \emph{Irradiation tests (IT):} Up to $5\,\%$ of dedicated test sensors and test structures are characterized after irradiation to ensure that the radiation effects on the delivered material do not change over production time.
\end{enumerate}
The information gathered from these quality control mechanisms is combined to judge the quality of the delivered sensor batches and decide whether the sensors will be integrated into the CMS detector.
In the following, this paper discusses process quality control.

\section{Test structure set}

	Process quality control relies on test structures. The structures are produced on the same wafers as the sensors, utilizing the space on the wafer edges surrounding the sensor. Because the test structures go through the same production process as the sensors, they exhibit the same properties and provide the means to investigate sensor and process parameters without having to measure the sensors directly. Specific test structures exist for each individual parameter. As a result of the specialized nature of the test structures, an individual measurement is generally quick. Thus, process quality control based on test structures achieves higher throughput rates than sensor measurements that strive to characterize every strip, pixel, or cell. Additionally, test structures provide access to parameters that cannot be measured directly on the sensors, including parameters that require potentially destructive measurements.

\begin{figure}[tb]
	\centering
	\includegraphics[width=\linewidth]{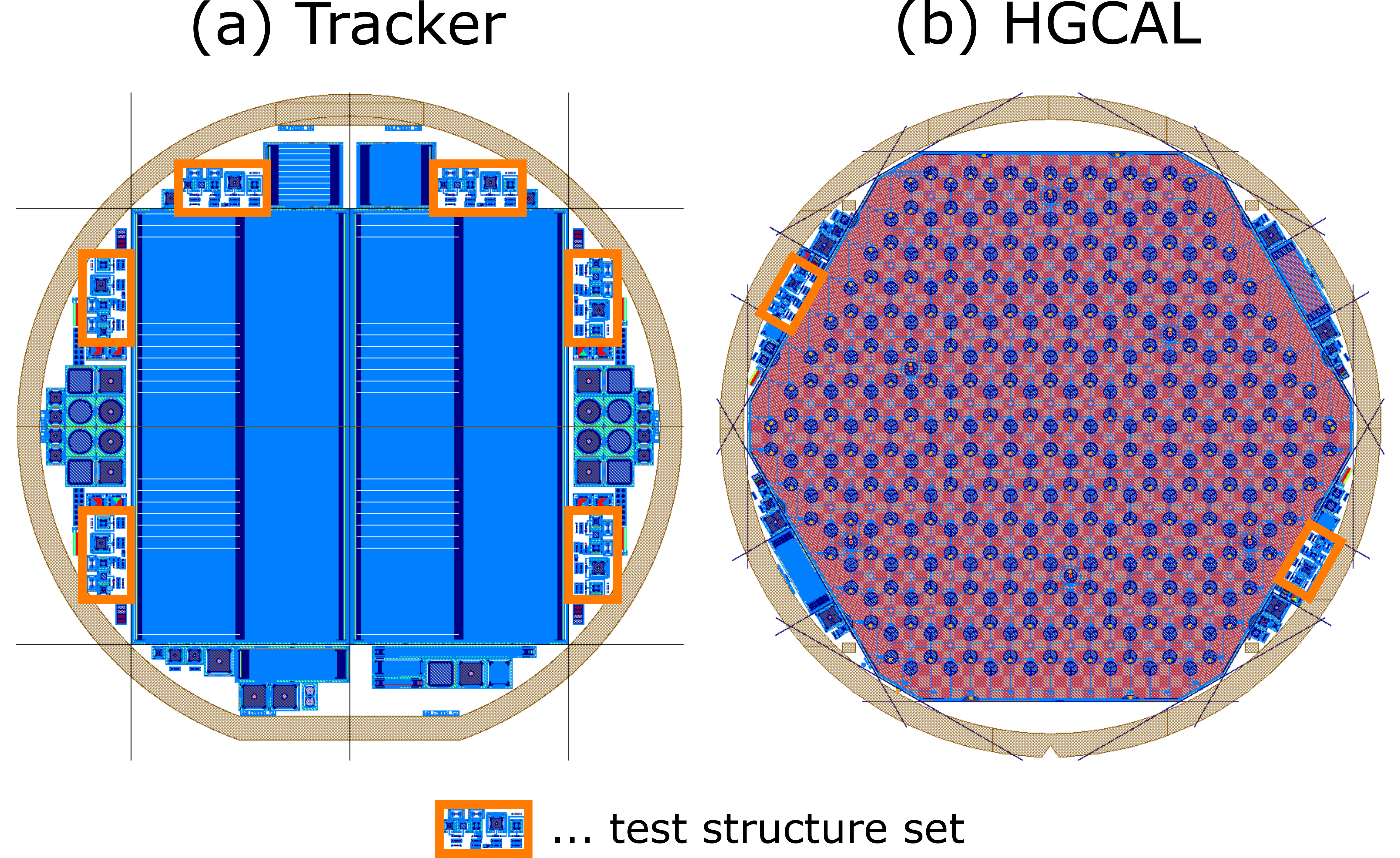}
	\caption{Layout schematic of an example tracker wafer (a) and a calorimeter wafer (b). Test structures are located in the periphery of the wafers. The standard set of test structures for CMS process quality control is highlighted on each wafer.}
	\label{fig:Wafers}
\end{figure}
	
	Figure~\ref{fig:Wafers} shows the layout of a tracker wafer with two PS-module strip sensors (Figure~\ref{fig:Wafers}a) and a \mbox{HGCAL} wafer (Figure~\ref{fig:Wafers}b). Various test structures and small test sensors are located in the periphery of the wafers. On every wafer, the set of test structures designed for automated process quality control is highlighted. The set is implemented six times on all tracker wafers and twice on all \mbox{HGCAL} wafers, which allows for tracking of process parameter variations across the wafer area. Because every wafer features the same design of the set, cross-comparability between production batches, test centers, and over the full three-year production time is maintained.
	
	\begin{figure*}[tb]
		\centering
		\includegraphics[width=\textwidth]{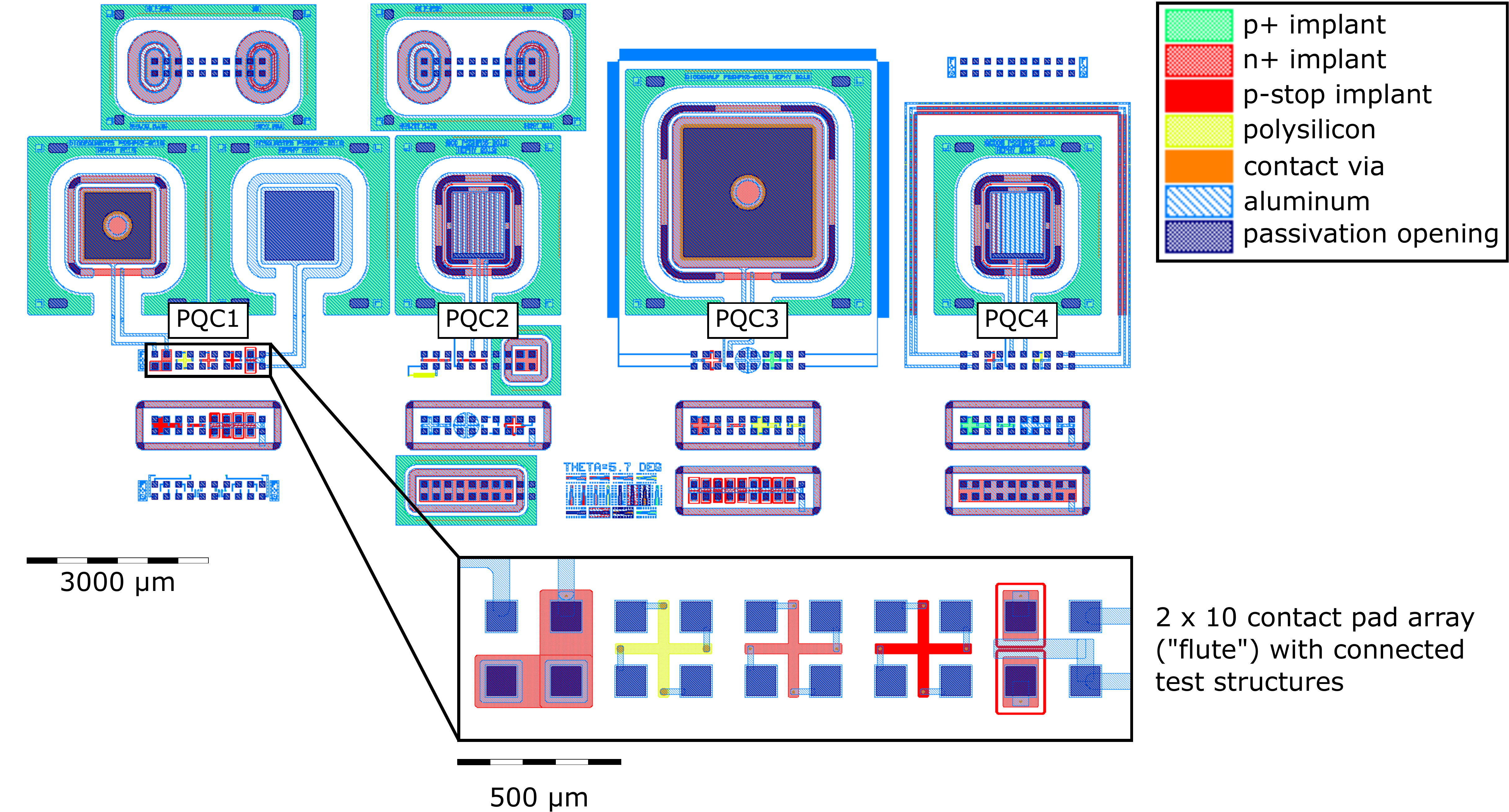}
		\caption{Test structure set for automated process quality control. The test structures are connected to arrays of 20 contact pads (a \enquote{flute}) that facilitate measurement using a 20-needle probe card. The four main flutes (\enquote{PQC1--PQC4}) provide a comprehensive overview of all relevant process parameters. A list of included test structures and process parameters is given in Table~\ref{tab:TSList}.}
		\label{fig:TSset}
	\end{figure*}
	
	A close-up of the set and its components is shown in Figure~\ref{fig:TSset}. The set facilitates automated measurements using a 20-needle probe card. For this purpose, all test structures are connected to contact pads structured into arrays, called \enquote{flutes}, of $2\times10$ pads each.
	
	\begin{table*}[tb]
		\caption{Overview of the four main flutes (\enquote{PQC1--PQC4}) of the test structure set (Figure~\ref{fig:TSset}), implemented test structures, and accessible process parameters.}
		\centering
		\begin{tabularx}{\textwidth}{ab}
			\toprule
			\textbf{Flute} & \textbf{\textit{Test structure} (process parameters)} \\
			\midrule
			PQC1 & \textit{1.56\,mm\textsuperscript{2} diode} (full depletion voltage, dark current, bulk resistivity, bulk doping concentration) \\
			& \emph{Metal-oxide-semiconductor capacitor} (flat band voltage, fixed oxide charge concentration, interface trap density, mobile oxide charges, oxide thickness) \\
			& \emph{Van-der-Pauw crosses} (\textit{n}\textsuperscript{+}, \textit{p}-stop, polysilicon sheet resistance) \\
			& \emph{Capacitors} (coupling capacitance, dielectric thickness) \\
			& \emph{Field-effect transistor} (threshold voltage, inter-strip properties) \\
			PQC2 & \emph{Gate-controlled diode with width ratio $\text{n\textsuperscript{+}}/\text{gate}=1$} (surface current, interface recombination velocity, interface trap density) \\
			& \emph{Polysilicon resistor} (bias resistance) \\
			& \emph{Line width structures} (\textit{n}\textsuperscript{+}, \textit{p}-stop implant line width) \\
			& \emph{Dielectric breakdown test structure} (dielectric strength of the coupling dielectric) \\
			PQC3 & \emph{6.25\,mm\textsuperscript{2} diode} (full depletion voltage, dark current, bulk resistivity, bulk doping concentration, correction for edge effects by combining results from both diodes of the set~\cite{Klanner}) \\
			& \emph{Bulk resistivity test structure} (bulk resistivity, bulk doping concentration) \\
			& \emph{Van-der-Pauw crosses} (metal, \textit{p}\textsuperscript{++} sheet resistance, \textit{p}\textsuperscript{++} implant line width) \\
			& \emph{Metal meander} (metal sheet resistance) \\
			PQC4 & \emph{Gate-controlled diode with width ratio $\text{n\textsuperscript{+}}/\text{gate}=1/3$} (surface current, interface recombination velocity, interface state density, generation lifetime~\cite{GCD}) \\
			& \emph{Cross-Bridge-Kelvin-Resistance test structures} (contact resistance metal to \textit{n}\textsuperscript{+} implant, metal to polysilicon) \\
			& \emph{Contact chains} (contact quality metal to \textit{n}\textsuperscript{+} implant, metal to \textit{p}\textsuperscript{++} implant, metal to polysilicon) \\
			\bottomrule
		\end{tabularx}
		\label{tab:TSList}
	\end{table*}
	
	The set provides access to all relevant process parameters (Table~\ref{tab:TSList}). It houses four main flutes (labeled \enquote{PQC1--PQC4}) that include test structures for each process parameter. The flutes \enquote{PQC1} and \enquote{PQC2} assess the most relevant parameters that provide a satisfactory overview of wafer properties. These parameters include the full depletion voltage, wafer resistivity, implant sheet resistances, fixed oxide charge concentration, quality of the coupling dielectric and inter-channel isolation, and Si-SiO\textsubscript{2} interface trap density. A measurement of both flutes is possible in about 30 minutes. The standard process quality control procedure foresees only the measurement of flutes \enquote{PQC1} and \enquote{PQC2} on two locations of each tested wafer.
	
	Flutes \enquote{PQC3} and \enquote{PQC4} complete the parameter set by providing access to additional sheet resistances, Si-SiO\textsubscript{2} interface generation lifetime, and contact quality. A measurement of all four flutes takes between one and two hours. A full characterization of all four flutes is performed at least once for every delivered wafer batch.
	
	In addition to flutes \enquote{PQC1--PQC4}, the set includes eleven flutes for in-depth analysis of individual process parameters in case of problems and an optical test structure to measure the alignment accuracy of the photolithography masks. Descriptions and measurement results that illustrate the functionality of selected test structures follow hereafter.
	
\subsection{MOS (Metal-Oxide-Semiconductor) capacitor}

MOS capacitors are standard tools for determining general properties of the Si-SiO\textsubscript{2} interface~\cite{Schroder}. They consist of an insulating SiO\textsubscript{2} layer sandwiched between the silicon bulk and a metal gate electrode. Fixed oxide charge $N_\text{ox}$, oxide trapped charge, mobile oxide charge, interface trapped charge, and oxide thickness $t_\text{ox}$ can be extracted from MOS capacitance--voltage ($C$--$V$) characteristics.

The CMS process quality control set contains a MOS capacitor with metal gate dimensions $1290\times1290\,\si{\micro\metre\squared}$ connected to the flute array labeled \enquote{PQC1}. High-frequency (i.\,e. $1$--$10\,\si{\kilo\hertz}$) $C$--$V$ sweeps are performed to extract the oxide flat-band voltage $V_\text{fb}$, from which the fixed oxide charge concentration $N_\text{ox}$ can be determined (assuming a negligible role of interface trapped charges) using
\begin{equation}
N_\text{ox}=\frac{C_\text{ox}/A\,(\phi_\text{ms}-V_\text{fb})}{q}\;.
\label{eq:MOS}
\end{equation}
Here, $C_\text{ox}/A$ denotes the oxide capacitance in accumulation divided by the gate area, $\phi_\text{ms}$ the metal--semiconductor work function difference, and $q$ the elementary charge.

\begin{figure}[tb]
	\centering
	\includegraphics[width=\linewidth]{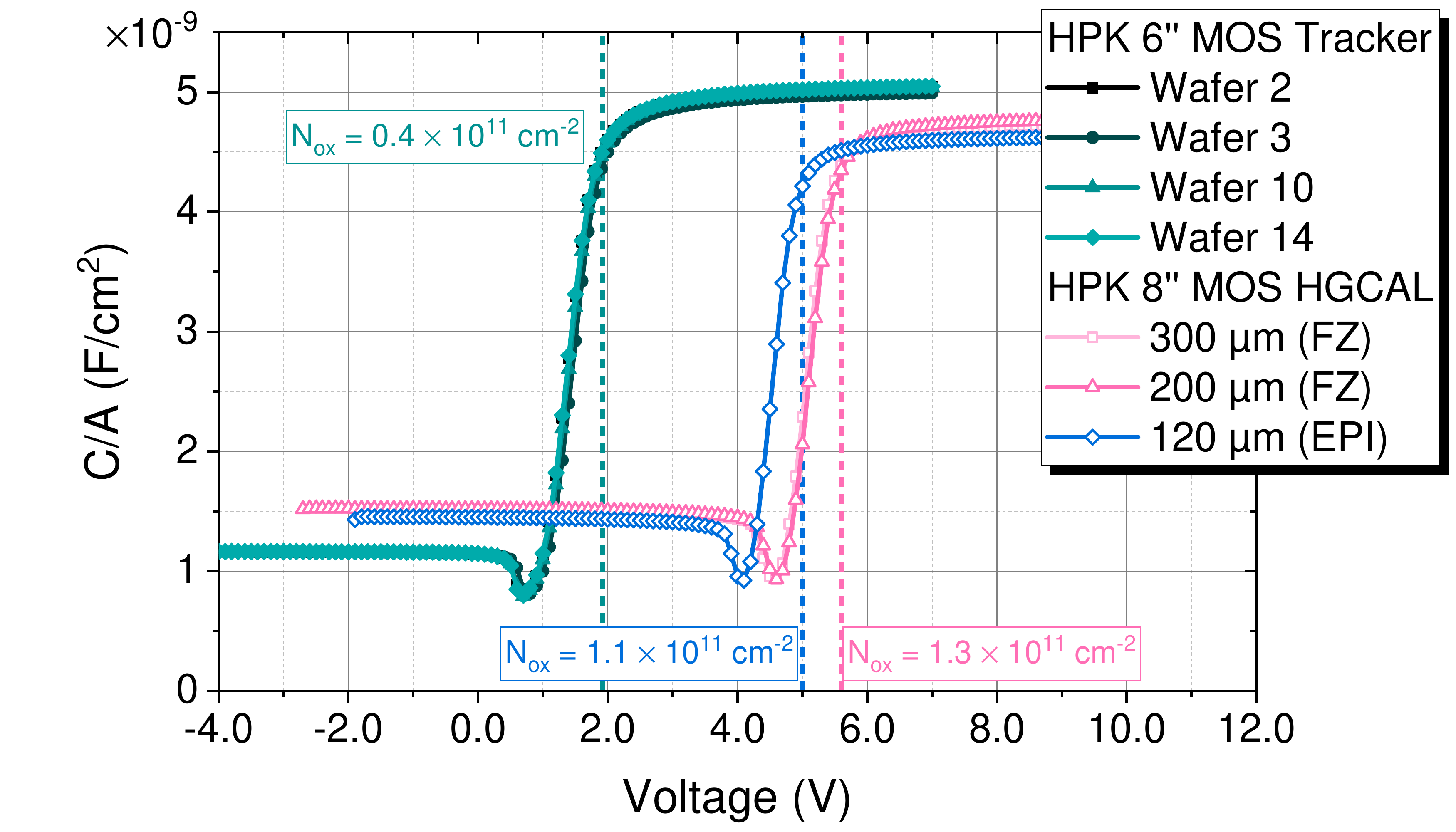}
	\caption{Comparison of MOS capacitor measurements on $6\,''$ tracker wafers and $8\,''$ \mbox{HGCAL} wafers. The fixed oxide charge concentration measured on \mbox{HGCAL} wafers is about three times higher than the value obtained for tracker wafers, which is attributed to the different production processes for tracker and \mbox{HGCAL} wafers. \mbox{HGCAL} wafers subjected to the standard float zone (FZ) process exhibit a $20\,\%$ higher fixed oxide charge concentration than those subjected to an epitaxial (EPI) process. The corresponding process parameters are summarized in Table~\ref{tab:MOS}.}
	\label{fig:MOS}
\end{figure}

\begin{table*}[tb]
	\caption{Comparison of process parameters extracted from MOS capacitor measurements on $6\,''$ tracker wafers and $8\,''$ HGCAL wafers, corresponding to Figure~\ref{fig:MOS}. HGCAL wafers are divided into float zone (FZ) and epitaxial (EPI) wafers.}
	\centering
	\begin{tabularx}{\textwidth}{lYYYY}
		\toprule
		Process & $C_\text{ox}/A$ ($\si{\nano\farad/\centi\metre\squared}$) & $V_\text{fb}$ ($\si{\volt}$) & $N_\text{ox}$ ($10^{11}\,\si{\per\centi\metre\cubed}$) & $t_\text{ox}$ ($\si{\nano\metre}$)\\
		\midrule
		Tracker ($6\,''$) & 5.0 & 1.9 & 0.4 & 705 \\
		HGCAL ($8\,''$ FZ) & 4.8 & 5.6 & 1.3 & 731 \\
		HGCAL ($8\,''$ EPI) & 4.6 & 5.0 & 1.1 & 748 \\
		\bottomrule
	\end{tabularx}
	\label{tab:MOS}
\end{table*}

$N_\text{ox}$ is found to vary substantially for different wafer production processes (Figure~\ref{fig:MOS}, Table~\ref{tab:MOS}). While the tracker wafers consistently exhibit a fixed oxide charge concentration of $N_\text{ox} = 4\times10^{10}\,\si{\per\centi\metre\squared}$, $N_\text{ox}$ is on average three times higher for \mbox{HGCAL} wafers. Furthermore, a difference between \mbox{HGCAL} float zone wafers and wafers subjected to an epitaxial process is observed. For the float zone process, $N_\text{ox} = 1.3\times10^{11}\,\si{\per\centi\metre\squared}$ is measured. The value found for epitaxial wafers ($N_\text{ox} = 1.1\times10^{11}\,\si{\per\centi\metre\squared}$) is about $20~\%$ smaller.

\subsection{Sheet resistance test structures}

\begin{figure}[tb]
	\centering
	\includegraphics[width=\linewidth]{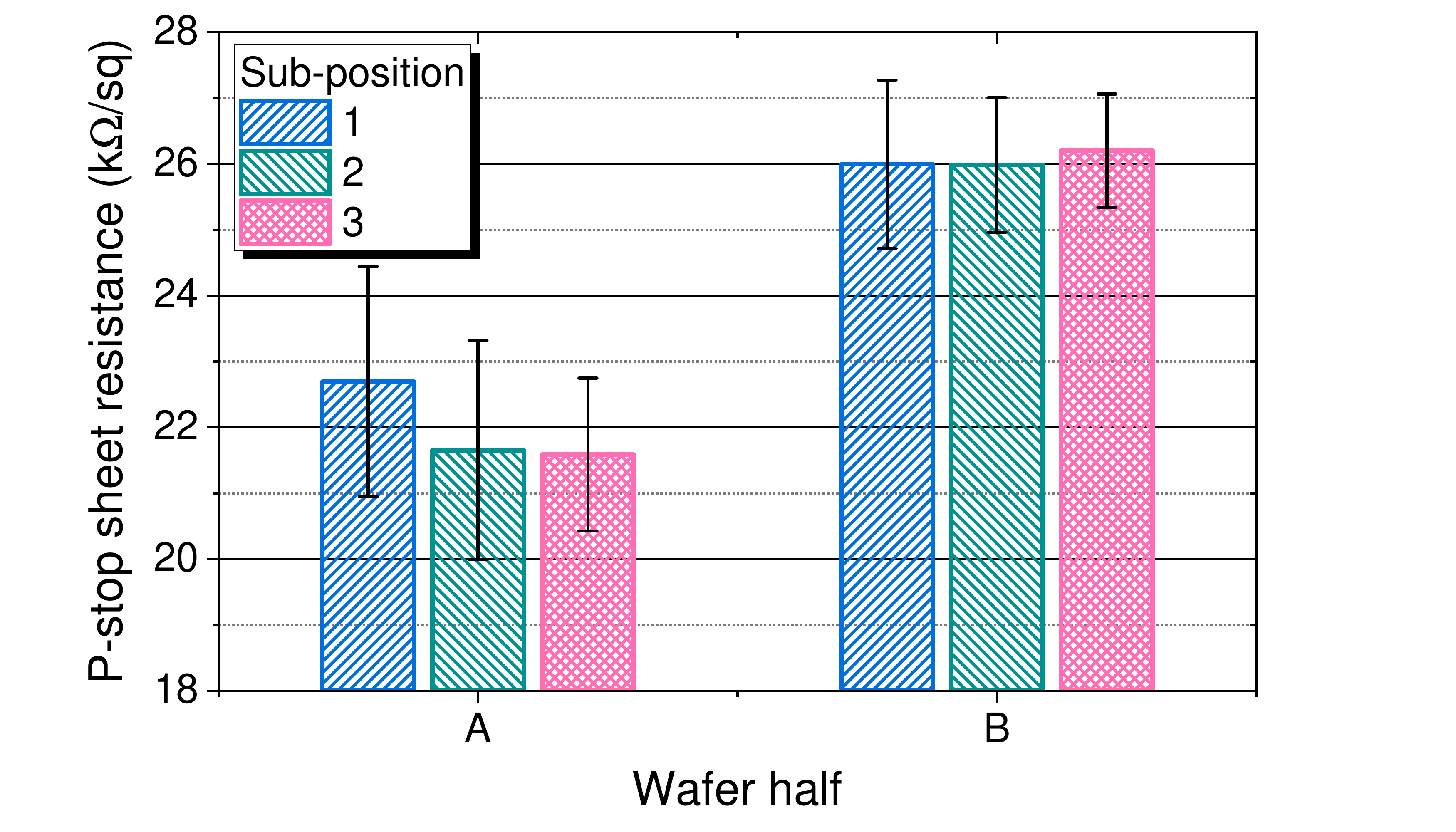}
	\caption{Sheet resistance of the \textit{p}-stop implant measured on Van-der-Pauw crosses at three sub-positions on both halves of the same tracker wafer. Half A exhibits lower sheet resistance values than half B. Consistent behaviour is found for the other wafers of the same production batch, indicating that the location dependency of the sheet resistance is an attribute of the production process.}
	\label{fig:VDP}
\end{figure}

Sheet resistance ($R_\text{sh}$) denotes the resistance of thin layers. It is related to the resistivity $\rho$ via the layer thickness $d$ by $R_\text{sh}=\rho/d$.

The CMS process quality control set includes Van-der-Pauw-type cross structures~\cite{Schroder,VDP} to determine the sheet resistance of \textit{n}\textsuperscript{+}, \textit{p}-stop, and \textit{p}\textsuperscript{++}, polysilicon, and aluminum layers. Additionally, elongated Van-der-Pauw cross bridges~\cite{Schroder} and meandering polysilicon and aluminum resistors are implemented to extract the respective layer width.

Measurements of \textit{p}-stop Van-der-Pauw crosses on tracker wafers find a dependency of the \textit{p}-stop sheet resistance on the location of the structure on the wafer (Figure~\ref{fig:VDP}). Each wafer is divided along the same axis into halves A and B; and, consistently for all measured wafers, half A exhibits lower sheet resistance values than half B ($R_\text{sh}\approx22\,\si{\kilo\ohm/\text{sq}}$ and $R_\text{sh}\approx26\,\si{\kilo\ohm/\text{sq}}$ respectively). These results indicate that production process specifics cause a location dependency of the \textit{p}-stop sheet resistance on the investigated samples.

\subsection{Field-effect transistors (MOSFETs)}

The CMS process quality control set utilizes \mbox{MOSFET} (Metal-Oxide-Semiconductor Field-Effect Transistor) test structures to investigate inter-channel properties. Especially the resistance between neighboring strips or cells (i.e. inter-channel resistance) is of interest. It is largely determined by inter-channel geometry, \textit{n}\textsuperscript{+}-implant distance, and \textit{p}-stop doping concentration and implantation depth. Before irradiation, the resulting resistance is typically on the order of $100\,\si{\giga\ohm}$, because of which direct measurement is subject to substantial errors.

\begin{figure}[tb]
	\centering
	\includegraphics[width=\linewidth]{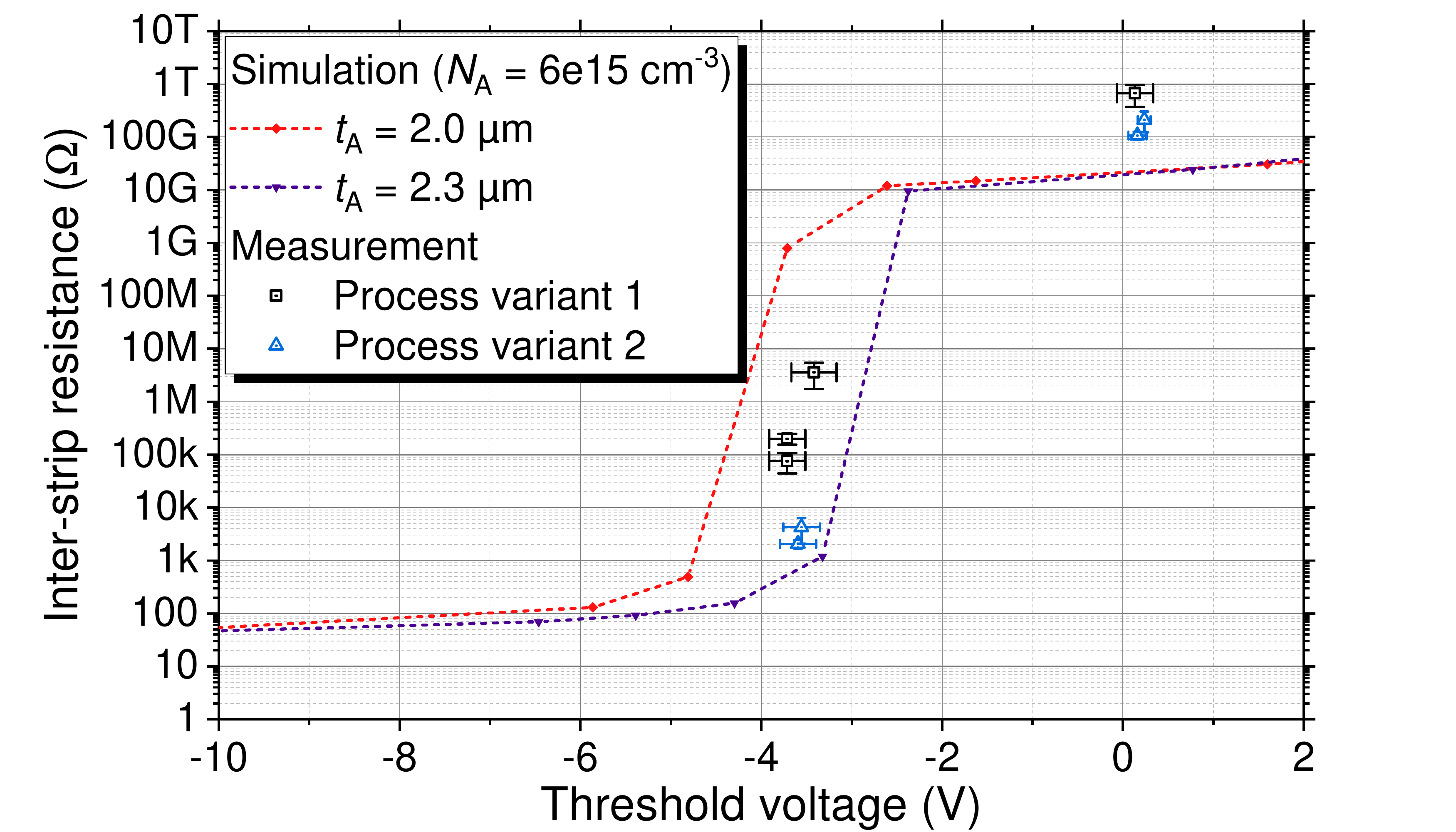}
	\caption{Sensor inter-strip resistance versus MOSFET threshold voltage as a result of varied \textit{n}-spray doping concentration. Measurement results for two process variants affecting dopant implantation depth (symbols) are compared to simulated characteristics for different \textit{p}-stop implantation depths $t_\text{A}$ (dotted lines). Simulation and measurement both observe a shift of the characteristics as a result of different \textit{p}-stop implantation depths. Figure adapted from~\cite{FETsHinger}.}
	\label{fig:FET}
\end{figure}

The MOSFETs implemented in the process quality control set allow tracking process variations that affect inter-channel properties without having to measure the inter-channel resistance directly.
The MOSFET inter-channel region replicates the sensor inter-channel layouts, including \textit{p}-stop implants below the gate between drain and source electrodes.
During process quality control, the MOSFET transfer characteristics are recorded, and the threshold voltage is extracted~\cite{MOSFET_Vth}.
The threshold voltage is sensitive to variations of \textit{p}-stop doping concentration and implantation depth~\cite{FETPaper} and relates to inter-channel resistance~\cite{FETsHinger} (Figure~\ref{fig:FET}).
Variations of the threshold voltage indicate changes of process specifics that affect inter-channel resistance.

\section{Conclusion}
\label{Sum}
	
 The CMS collaboration has set up a dedicated quality assurance plan for the series production of silicon sensors for the upcoming Phase-2 Upgrade of the CMS Outer Tracker and Endcap Calorimeter. The strategy foresees three main quality control mechanisms: sensor quality control, process quality control, and irradiation tests.
 Process quality control will utilize a test structure set that provides access to all relevant process parameters. The set is optimized for automated measurements using a 20-needle probe card.
 Preliminary measurements on prototypes have demonstrated the functionality of the test structures included in the set.
 Six instances of the set will be implemented on all tracker wafers and two instances on all \mbox{HGCAL} wafers. During the production period -- scheduled between 2020 and 2023 -- about $20\,\%$ of each delivered wafer batch will be subjected to process quality control.

\section*{Acknowledgements}
\label{thanks}
	The research leading to these results received funds from the call "Forschungspartnerschaften"
	of the Austrian Research Promotion Agency (FFG) under the grant no. 860401.

%% The Appendices part is started with the command \appendix;
%% appendix sections are then done as normal sections
%% \appendix

%% \section{}
%% \label{}

%% If you have bibdatabase file and want bibtex to generate the
%% bibitems, please use
%%
%% \bibliographystyle{elsarticle-num}
%% \bibliography{mybib}

%% else use the following coding to input the bibitems directly in the
%% TeX file.

\end{document}